\documentclass{vgtc}                          % final (conference style)
\graphicspath{{figures/}{pictures/}{images/}{./}} % where to search for the images

\usepackage{times}                     % we use Times as the main font
\usepackage{tabu}                      % only used for the table example
\usepackage{array}
\usepackage{diagbox}
\usepackage{booktabs}                  % only used for the table example
\usepackage{lipsum}                    % used to generate placeholder text
\usepackage{mwe}                       % used to generate placeholder figures
\DeclareUnicodeCharacter{0301}{\'{e}}

\usepackage{mathptmx}                  % use matching math font

\onlineid{0}

\vgtccategory{Research}

\vgtcinsertpkg

\title{ViSTooth: A Visualization Framework for Tooth Segmentation on Panoramic Radiograph}

\author{Shenji Zhu\thanks{e-mail: 231330018@hdu.edu.cn}\\ %
        \scriptsize Hangzhou Dianzi University %
\and Miaoxin Hu\thanks{e-mail: miaoxinhu@outlook.com}\\ %
     \scriptsize Hangzhou Dianzi University %
\and Tianya Pan\thanks{e-mail: cbpantianya@163.com}\\ %
     \scriptsize Hangzhou Dianzi University %
\and Yue Hong\thanks{e-mail: hy984500@163.com}\\ %
     \scriptsize Department of Stomatology, First Affiliated Hospital, Zhejiang University. \\ \scriptsize School of Medicine, Zhejiang University. %
\and Bin Li\thanks{e-mail: libin330624@163.com}\\ %
     \scriptsize Department of Stomatology, Shengzhou People's Hospital. %
\and Zhiguang Zhou\thanks{e-mail: zhgzhou@hdu.edu.cn}\\ %
    \scriptsize Hangzhou Dianzi University
\and Ting Xu\thanks{e-mail: xut@zju.edu.cn}\\ %
     \scriptsize Department of Stomatology, First Affiliated
Hospital, Zhejiang University. %
     }

\teaser{
  \centering
  \includegraphics[width=\linewidth]{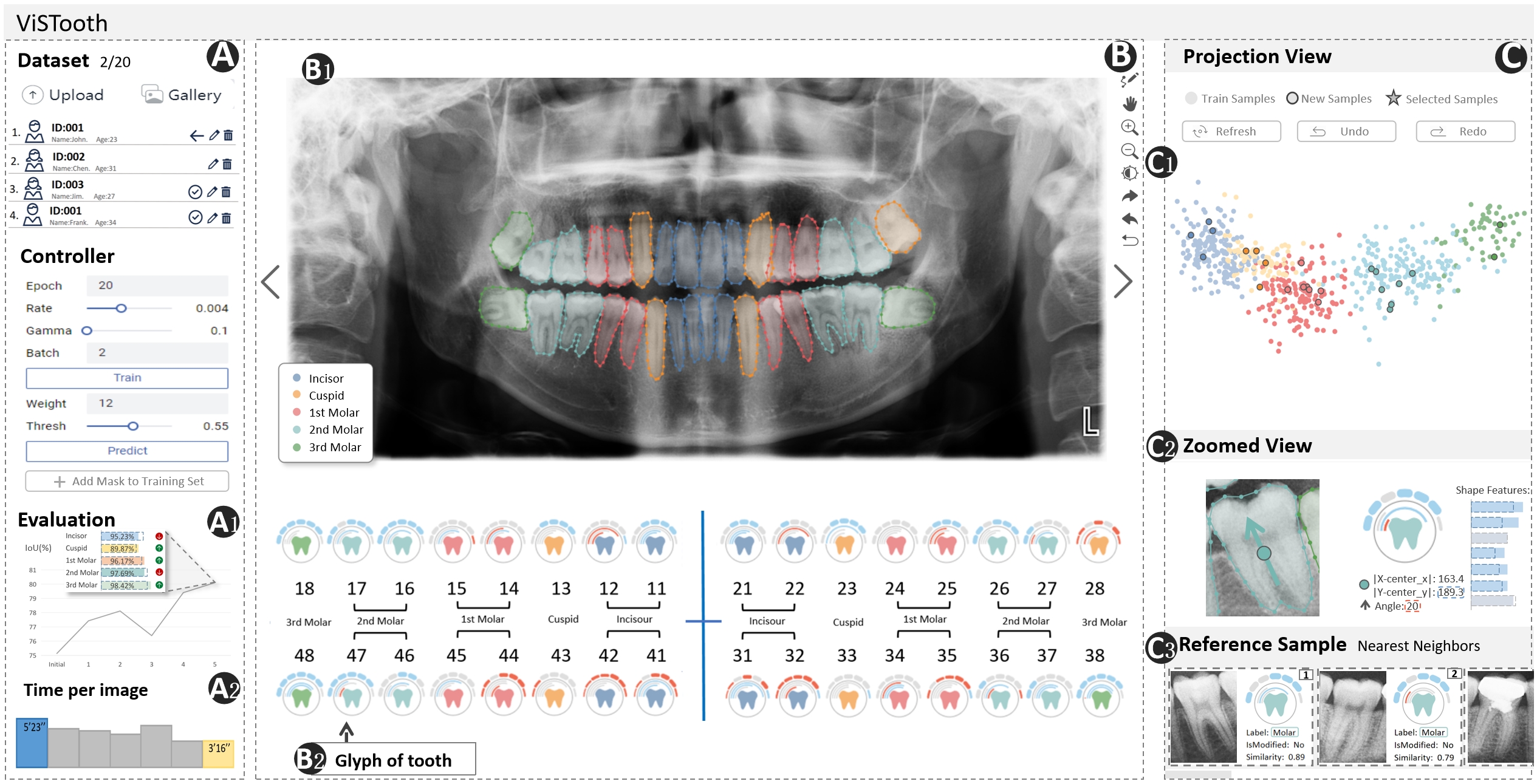}
  \caption{The visualization interface for tooth segmentation on panoramic radiograph. (A)Model Explorer: The dataset panel displays the basic information of the unlabeled panoramic radiograph and the control panel controls the values of parameters in the model such as training times, learning rate and so on. (A1)The line chart to display the process of model optimization. (A2)The barchart to show the time of manually correction per image. (B)Radiographic Feature Explorer: (B1)The Panoramic View to show the segmentation masks on panoramic radiograph. (B2)The Glyph View to reveal the features of tooth segmentation. (C)Extracted Feature Explorer: (C1)The Scatterplot View to provide an overview of relationships among the tooth samples. (C2)The Zoomed View to display a more detailed exploration of extracted features. (C3)The Reference Sample View to illustrate the attributes of similar instances.}
  \label{fig:teaser}
}

\abstract{
    Tooth segmentation is a key step for computer aided diagnosis of dental diseases. Numerous machine learning models have been employed for tooth segmentation on dental panoramic radiograph. However, it is a difficult task to achieve accurate tooth segmentation due to complex tooth shapes, diverse tooth categories and incomplete sample set for machine learning. In this paper, we propose ViSTooth, a visualization framework for tooth segmentation on dental panoramic radiograph. First, we employ Mask R-CNN to conduct preliminary tooth segmentation, and a set of domain metrics are proposed to estimate the accuracy of the segmented teeth, including tooth shape, tooth position and tooth angle. Then, we represent the teeth with high-dimensional vectors and visualize their distribution in a low-dimensional space, in which experts can easily observe those teeth with specific metrics. Further, we expand the sample set with the expert-specified teeth and train the tooth segmentation model iteratively. Finally, we conduct case study and expert study to demonstrate the effectiveness and usability of our ViSTooth, in aiding experts to implement accurate tooth segmentation guided by expert knowledge.
} % end of abstract


\begin{document}

%% The ``\maketitle'' command must be the first command after the
%% ``\begin{document}'' command. It prepares and prints the title block.

%% the only exception to this rule is the \firstsection command
\firstsection{Introduction}

\maketitle
%% \section{Introduction} %for journal use above \firstsection{..} instead
% This template is for papers of VGTC-sponsored conferences which are \emph{\textbf{not}} published in a special issue of TVCG.
Panoramic radiograph is a widely-used imaging modality for dental examination in stomatology, which provides a visual representation of all teeth within the dental cavity, and helps doctors to examine the pathological conditions, such as dental calculus, dental malformations and caries\cite{jcm9072313}. Tooth segmentation is a pivotal step for computer aided diagnosis of tooth-related disorders\cite{diagnostics14050497}. However, manual annotation is a laborious and time-consuming task, especially when there are overlapping shadows or low contrast.

In recent years, numerous machine learning models have been proposed for tooth segmentation on dental panoramic radiograph\cite{2018Automated,ACompre2023,Develop2023}, encompassing both unsupervised and supervised methods. Unsupervised methods include threshold-based segmentation\cite{Indraswari2015, Mohamed2014}, edge detection\cite{razali2014, 2012fuzzy}, and graph theory\cite{li2022semantic}, while supervised methods rely on labeled data for training\cite{SILVA201815}. Deep learning-based approaches, such as U-Net\cite{Unet2023,Koch2019}, Faster R-CNN\cite{chen2019deep,kim2020automatic} and PANet\cite{2020study}, fall under the category of supervised methods. However, due to the substantial variations in tooth shape and types, such a strategy is still unable to fundamentally solve the problem of accuracy and robustness of automatic segmentation algorithms, which brings uncertainty to the subsequent diagnosis\cite{Schwendicke2020, SILVA201815, leite2021artificial}.

Extensive discussions with professional dental experts and computer experts have led to the consensus that traditional AI segmentation methods exhibit significant limitations when applied to tooth segmentation on panoramic radiograph. Two primary questions have been identified. \textbf{Q1.} The model has a primary focus on pixel features but lacks essential dental expertise like tooth shape and angle, hindering its contextual understanding and ability to discern intricate details and nuances specific to dental conditions. This limitation becomes particularly evident when the training sample is unable to cover the full spectrum of dental conditions, causing the model to struggle in accurately segmenting complex or special teeth. \textbf{Q2.} In the process of segmentation, complete dependence on automated algorithms may lead to suboptimal tooth segmentation results in certain cases, as the model lacks the ability to dynamically adapt and refine its segmentation outputs in response to the nuances of individual cases.

As such, we develop a human-machine collaboration framework, VisTooth(Figure \ref{fig:teaser}), comprehensively considering tooth features and introducing expertise in the segmentation process to optimize the model outcomes. Firstly, we use the fine-adjusted Mask R-CNN model\cite{maskrcnn2017} to achieve preliminary tooth segmentation, which focuses on the features involving dental expertise. A glyph representation is generated to visualize these features(Q1). For cases such as structural abnormalities or blurred tooth contours, human expert judgment and intervention are needed to improve segmentation quality\cite{DBLP2021}. To address this, we develop an interactive tool to allow experts to correct the outcomes of the initial segmentation. Further, different levels of detail information views are incorporated to assist experts in screening out high-quality segmentation , and expert-specified teeth will be piped into the model for interactive optimization (Q2). Finally, in order to demonstrate the effectiveness and usability of our VisTooth in addressing the issue of tooth segmentation, we conduct case study and expert study.

The major contributions of this paper are listed as follows:
\begin{itemize}
\setlength{\itemsep}{2pt}
\setlength{\parsep}{2pt}
\setlength{\parskip}{2pt}
\item[$\bullet$] A set of feature metrics are proposed to assess the segmentation according to dental expertise.
\item[$\bullet$] A novel visual analytics system is implemented to summarize and compare the tooth segmentation with different levels of details.
\item[$\bullet$] A new human-machine collaboration workflow leveraging advanced machine learning algorithms and human expertise is implemented to guarantee the accuracy and efficiency of tooth segmentation.
\end{itemize}

\section{RELATED WORK}
\subsection{Tooth Segmentation}
Tooth segmentation on panoramic radiograph is a critical task, addressed through two primary methodologies: unsupervised and supervised approaches. In the unsupervised category, various strategies have been developed. Modi et al.\cite{Modi2011} proposed a region-based method to identify regions of interest for gap valley and tooth isolation using binary edge intensity integral curves. Indraswari et al.\cite{Indraswari2015} employed a three-step process involving directional image formation using DDFBT, enhancement for edge reinforcement and noise removal, and MAT with Sauvola Local Thresholding for segmentation. Alsmad et al.\cite{Alsmadi2018} utilized a cluster-based approach, while Hasan et al.\cite{Hasan2016} focused on jaw segmentation using gradient information in a four-step method comprising k-means clustering, point detection around the jaw, gradient vector flow snakes, and shape correction for the segmented area. Li et al.\cite{Li2012} introduced a new watershed algorithm based on mathematical morphology, specifically tailored for dental X-ray image segmentation. Fariza et al.\cite{Fariza2019} employed a method to extract different dental structures using conditional spatial fuzzy C-means clustering.

In contrast, supervised methods leverage deep learning models trained on annotated data to improve segmentation accuracy and stability. Jader et al. \cite{jader2018} are credited for being the pioneers who detected and segmented each tooth on panoramic radiographs. Almalki et al.\cite{ALmalki2023} applied two self-supervised learning methods to Swin Transformer on dental panoramic radiographs: SimMIM and UM-MAE. Zhang et al.\cite{Zhang2018AnET} proposed a novel method that using label tree with cascade network structure combining several key strategies for teeth recognition, which can deal with many complex cases. Helli et al.\cite{helli2022} employed a two-step method where they employed a U-Net to create prediction followed by post-processing operations to achieve segmentation. Leite et al.\cite{leite2021artificial} proposed a CNN-based solution for determining tooth contours using semantic segmentation, further refined by a Fully Convolutional Network (FCN). These methods were evaluated using metrics such as average Intersection over Union (IoU) and Hausdorff distance, compared against manual annotations and medical software. Tuzoff et al.\cite{tuzoff2019tooth} applied the Faster R-CNN object detection model to generate tooth borders, enhancing the output through integration with the VGG16 classification network and heuristic rules of dentition arrangement. Additionally, Mask R-CNN\cite{maskrcnn2017}, a deep learning-based method, offers simultaneous object detection and segmentation, incorporating ROI Align for improved accuracy. Despite the higher precision of supervised methods, challenges persist in scenarios of insufficient or inaccurately annotated training data, underscoring the ongoing need for accurate segmentation on panoramic radiograph\cite{bhat2023}. 

In this paper, we select appropriate neural network and incorporate a consideration for dental expertise when using Mask R-CNN for tooth segmentation. Further, we prpose a novel human-computer interaction system that allows experts to interactively refine segmentation outputs, thereby enhancing the accuracy and reliability of the final results.

\subsection{Visualization for Artificial Intelligence}
In the domain of artificial intelligence, the burgeoning complexity of models necessitates advanced methods for elucidation of their inner workings. Visualization tools play an instrumental role in this context, aiding in the comprehension of training data, model architecture, and output\cite{VisAI2023, DAI2018}. A notable contribution in this field is the OoDAnalyzer\cite{Ood} by Chen et al., which presents an interactive visual method for the identification and explanation of Out-of-Distribution (OoD) samples. Kandel et al.\cite{kandel2012} proposed Profiler, a tool designed to assess quality issues in tabular data. Anomaly detection methods are employed to detect and categorized data anomalies. And visual summaries aids evaluation of potential anomalies and their causes. Liu et al.\cite{liu2017analyzing} developed a visual analytics approach using time series data to represent training dynamics of Deep Generative Models (DGMs). It includes a novel blue noise line sampling scheme and a credit assignment algorithm for improved understanding and diagnosis of DGM training processes. Cao et al.\cite{cao2020analyzing} presented a visual analysis tool AEVis to explain why adversarial examples are misclassified. The contribution analysis and rich interactions further enable users to trace the root cause of the misclassification of adversarial examples. Wang et al.\cite{wang2021} presented CNN EXPLAINER, an interactive visualization tool designed for non-experts to learn and examine convolutional neural networks. Through smooth transitions across levels of abstraction, users can inspect the interplay between operations and outcomes. Mahendran et al.\cite{Ma2015} introduced the Deep Visualization Toolbox (DeepVis) to visualize and interpret CNN features by synthesizing input images that maximally activate specific neurons. Selvaraju et al.\cite{Sel2017} introduced Grad-CAM, which has since been widely adopted for interpreting CNN-based models in various domains, including medical imaging and natural language processing. Chen et al.\cite{chen2023unified} introduced Uni-Evaluator, an open-source visual analytic tool for model evaluation tasks like target detection. It represents predictions as probability distributions across tasks, using matrices, tables, and grids for comprehensive evaluation from a global to sample level. Humans can also monitor the learning process and evaluate the effectiveness of AI models at any time through visualization\cite{VisAI2023}. Ahn et al.\cite{ahn2019fairsight} proposed a visual analytic system FairSight to capture both the global and instancelevel fairness with evidence of potential unfair outcomes. 

Collectively, these developments underscore the pivotal role of visual analytics in the interpretation, evaluation, and refinement of complex AI models within the scientific community. In contrast, we apply interactive visual analytics to the detection and correction of mask errors in the process of automatic segmentation, aiming to facilitate the high-quality of outputs.

\section{TASK ANALYSIS AND SYSTEM OVERVIEW}
In this section, we provide a summary of analysis tasks(T1-T4) identified through interviews with domain experts and subsequently present the pipeline of the proposed visual analysis system.

\subsection{Task Analysis}
Our system was developed through a collaborative effort involving experts in dental examination (E1 and E2) and an expert in graphics and visualization (E3). E1 and E2 are highly experienced oral and maxillofacial radiologists each possessing over 5 years of extensive expertise. E3 is a seasoned professor specializing in data visual analysis. In the early stages of our collaboration, weekly meetings were conducted with these three experts to seek opportunities to optimize the process of tooth segmentation through literature review. According to experts, the diversity of teeth presents significant challenges to current AutoML approaches. To ensure the accuracy of tooth segmentation, further expert judgment and correction are deemed necessary. Consequently, we delved into the design requirements of a human-machine collaborate system. From these discussions, we derived four key analytical tasks, summarized as follows:

\textbf{T1. Integration of dental expertise into the segmentation process.}
General segmentation method only considers the pixel features\cite{pixelfeature2023}. However, dental expertise like the regularity in the physiological structure and arrangement characteristics of teeth can provide valuable information for segmentation. Hence, the employed feature extraction network of model should be adept at identifying the intricate structures on panoramic images\cite{bhat2023}. And the workflow should also incorporate a consideration for arrangement features when determining tooth labels.  

\textbf{T2. Assessment of Automatic Tooth Segmentation Accuracy.}
Once obtaining preliminary automated outputs, the subsequent step is to exam and modify potentially incorrect segmentation masks. In order to facilitate this correction process, it is necessary to propose quantifiable metrics to assess the accuracy of automatic tooth segmentation results. Additionally, a clear visual cue should also be displayed to guide the experts to review and manually correct.

\textbf{T3. Model Optimization through Valuable Instance Sampling.}
The diversity in different types of teeth presents a challenge for the machine learning model. However, manual corrections capture expert expertise on the accurate delineation of tooth boundaries which can serve as high-quality labeled data to retrain the ML-model. Thus, it becomes necessary to incorporate valuable expert corrections as a complement to the training set to optimize the model especially when the initial sample set is hard to encompass all possible variations.

\textbf{T4. Development of an Interactive Tooth Segmentation Tool.}
To ensure the effectiveness of subsequent work, it is deemed crucial to develop an interactive tool that can implement accurate tooth segmentation and with continuous iterative optimization. To the best of our knowledge, our work is the first attempt to provide a combination of man-machine tooth segmentation tool.

\subsection{System Overview}
\begin{figure*}
\centering 
\includegraphics[width=1\linewidth]{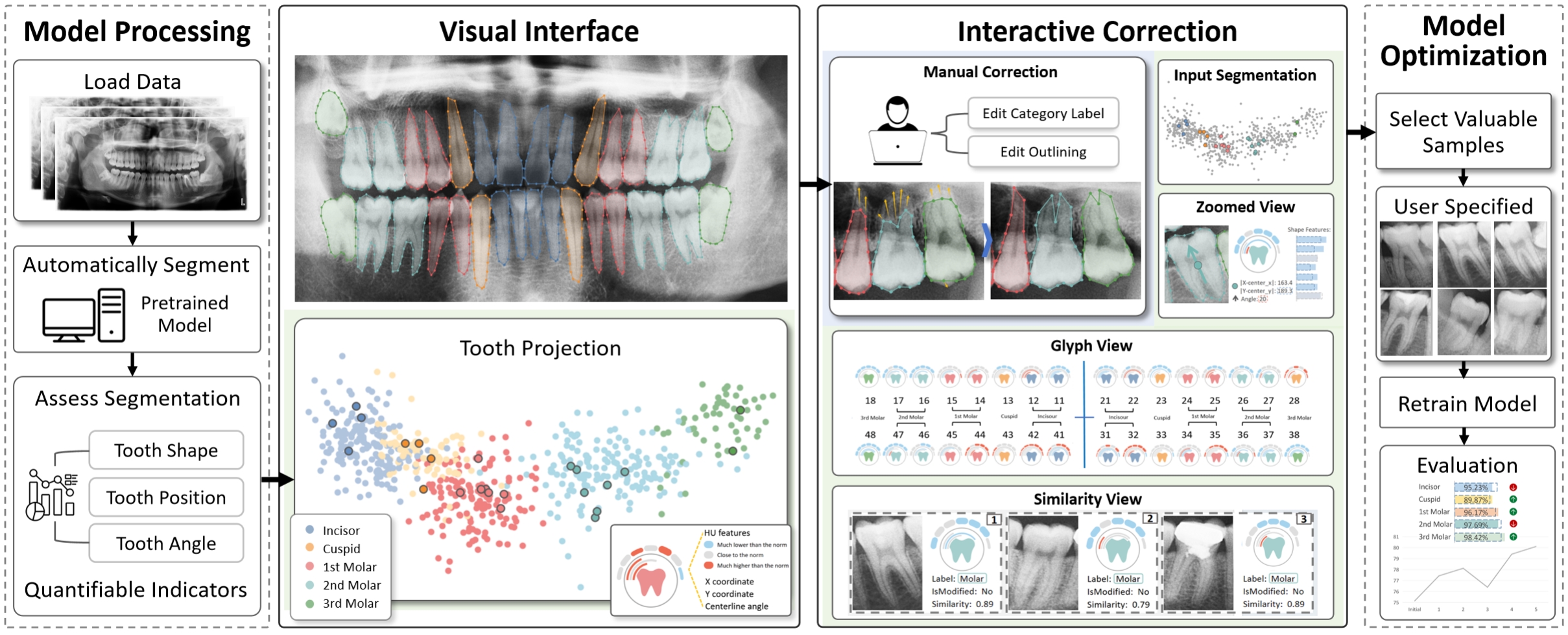}
\caption{The pipeline of VisTooth for tooth segmentation on panoramic radiograph.} %最终文档中希望显示的图片标题
\label{pipeline3} %用于文内引用的标签
\end{figure*}

Motivated by the identified tasks, we propose a visualization framework enabling experts to efficiently achieve high-quality tooth segmentation on panoramic radiograph. The system pipeline is depicted in Figure \ref{pipeline3}. Initially, the Mask R-CNN model is trained with a certain amount of manual labeled data, categorizing teeth into five classes: incisor, canine, 1st, 2nd, and 3rd molar. To estimate the outputs of the model, we propose several quantifiable metrics including tooth shape, tooth position and tooth angle. Concurrently, we devise a glyph-based visualization scheme to represent these information, thereby offering experts a comprehensive set of evaluation criteria(T1). We develop a scatterplot view to provide an overview of relationships among the tooth samples, so that the possible inaccurate results can be identified by the abnormal distribution(T2). We provide experts with visual interface to show the initial segmentation of the model and interactive tools for error correction. Then the corrected high-quality tooth samples, selected by experts, are fed back into the model for adaptive iterative optimization(T3). Ultimately, a human-machine collaborative visual tool is developed for the segmentation of teeth(T4).

\section{VISTOOTH}
We propose a visualization framework, ViSTooth, that integrates automatic technologies and interactive visualization to support human-machine collaboration for accurate tooth segmentation. This section introduces four key components: data labeling, tooth segmentation model, visualization design and model optimization.

\subsection{ Data Labeling}
The panoramic radiographs used in this study were selected from a patient image database at the hospital. The patients gave their informed consent before any panoramic radiographs were taken, and their privacy was protected when using the data for medical research. The dataset comprises 521 panoramic radiographs. We selected 300 images for experts to mark ground truth segmentation labels randomly, while the remaining 221 images were used as a test set. This process was under a supervision of two dentists(E1 and E2) using a tagging tool developed with the Python programming language. We attended weekly meetings where related issues were discussed and the labels were reviewed to assure quality. In the end, the 300 labeled images with ground truth segmentation labels was divided into a training set(240 images) and a validation set(60images). The study was approved by the Ethics Committee of The First Affiliated Hospital, Zhejiang University School of Medicine. (approval no. 20230785)

\subsection{ Tooth Segmentation Model}
\begin{figure} 
\includegraphics[width=1\linewidth]{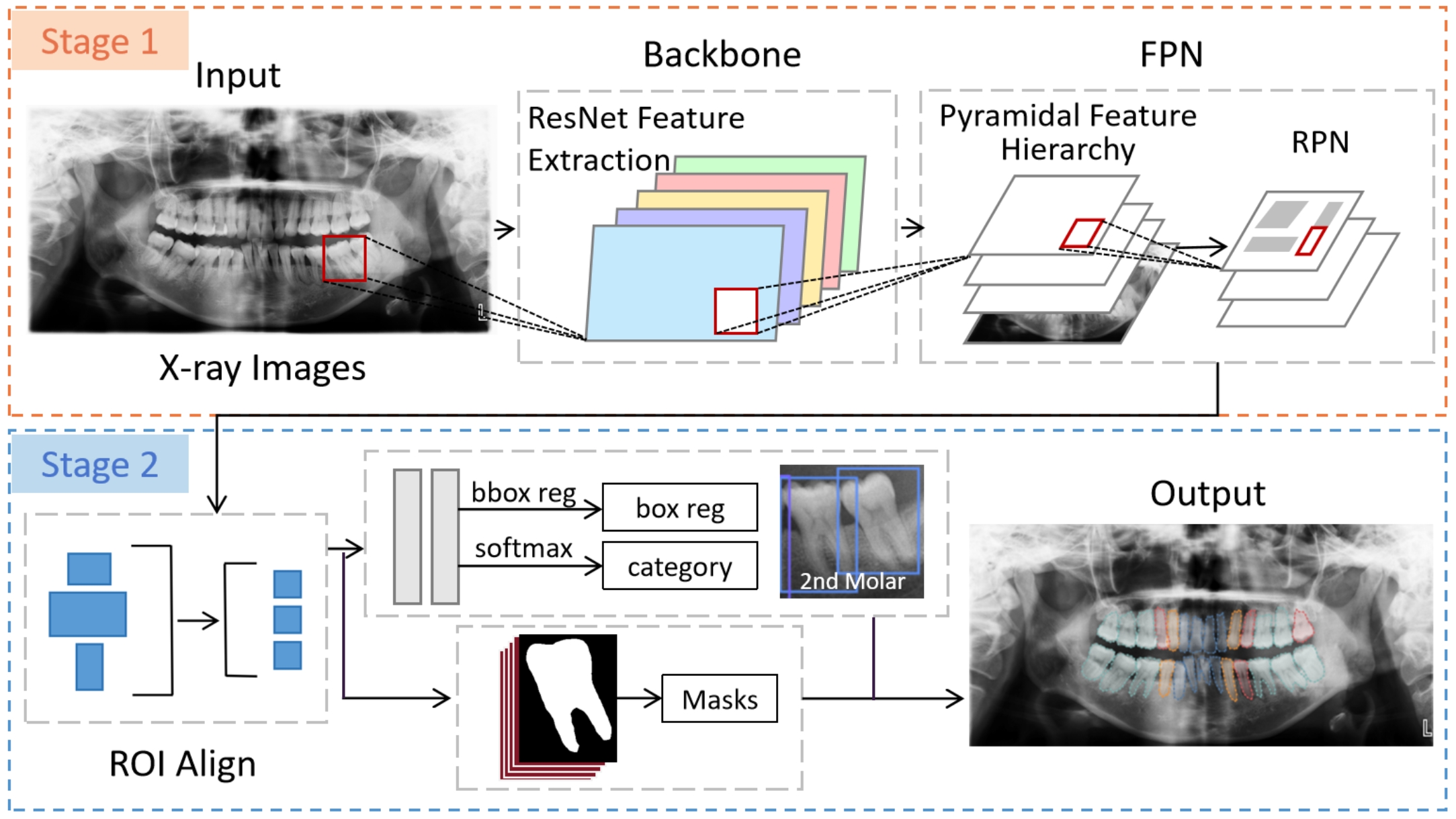}
\caption{The illustration of Mask R-CNN.} %最终文档中希望显示的图片标题
\label{illustration} %用于文内引用的标签
\end{figure}

In this paper, we emloy the Mask R-CNN model for teeth segmentation on panoramic radiograph. Mask R-CNN is a two-stage instance segmentation framework, as depicted in Figure \ref{illustration}. Specifically, the first stage proposes candidate tooth bounding boxes regardless of categories. Fistly, the panoramic radiograph is fed into the backbone to extract features. Then the features compose a pyramid network (FPN) to generate candidate regions with the potential to contain tooth structures. Since Mask R-CNN is a flexible framework, we tried to change the feature extraction network in backbone to make the model more suitable for panoramic segmentation tasks, including ResNet networks with 50, 101 and 152 layers\cite{he2016deep} and VGG16 network\cite{VGG2014}. As shown in Table \ref{table1}, We find that ResNet50 is the optimal choice for panoramic radiograph due to its fewer layers, which can refrain from overfitting, and its overall IoU score reaches 75.14\%.

\begin{table*}
\centering
\caption{Comparison of evaluation metrics with different backbones.}
\label{table1}
\renewcommand\arraystretch{1.65}
\begin{tabular}{|c|c|c|c|c|c|} 
\hline
Model                        & Backbone                    & IoU(\%)                & Precision(\%)          & Recall(\%)             & F1-score(\%)            \\ 
\hline
\textbf{\textbf{Mask R-CNN}} & \textbf{\textbf{ResNet-50}} & \textbf{\textbf{75.1}} & \textbf{\textbf{75.7}} & \textbf{\textbf{83.5}} & \textbf{\textbf{79.4}}  \\ 
\hline
Mask R-CNN                   & ResNet-101                  & 65.3                   & 65.9                   & 73.7                   & 69.6                    \\ 
\hline
Mask R-CNN                   & ResNet-152                  & 53.4                   & 53.9                   & 58.1                   & 55.9                    \\ 
\hline
Mask R-CNN                   & VGG16                       & 71.2                   & 71.8                   & 81.1                   & 76.2                    \\
\hline
\end{tabular}
\end{table*}

The second stage is termed as the R-CNN stage, which extracts features using RoIAlign\cite{maskrcnn2017} for each proposal and performs proposal classification, bounding box regression and mask predicting. This involves corresponding each pixel on the original panoramic radiograph with the feature map and matching it with preset fixed features. Subsequently, the model conducts multi-classification on these candidate regions, generating masks to complete the segmentation task. During the training stage, we classified sample teeth into five categories: incisors, cuspids, 1st and 2nd molars, and 3rd molar. In the process of classification, we guide the model to not only consider the image features of the segmented targets but also introduce heuristic rules based on the order of tooth arrangement. When image features are blurred and difficult to discern, priority is given to the segmentation category determined by the arrangement order.

\subsection{Visualization for Tooth Segmentation}
Due to the above AutoML segmentaion approach not always being accurate, in this section, we design the visualization interface to present the segmentation results from the model and to support more detailed feature exploration. Figure \ref{fig:teaser}, displays the visual interface of our system, which comprises a control panel and five maim views.

\subsubsection{Segmentation Explorer Component}
The radiographic feature exploration component contains two sub-views: a panoramic view and a glyph view. 

As shown in Figure \ref{fig:teaser}(B1), the panoramic view visualizes the tooth segmentation outcomes generated by the system. It facilitates a direct comparative analysis for experts to assess the congruence between the segmented contours and ground truth. Experts can adjust the initial segmentation mask by dragging contour points, ensuring a closer alignment with the actual targets.

The glyph view represents the detailed features of segmentation outputs. In this paper, we propose three essential metrics of tooth segmentation including shape, coordinates and center-line angle. Subsequently, we employ a visual prompting approach to guide experts in making more nuanced judgments and corrections to these results.

Firstly, we use the HU moment\cite{hu1962visual, HUap2020} as the shape feature of the segmentation mask to characterize individual teeth, which is calculated as follows:

\begin{equation}
m_{p,q}=\sum_{x}\sum_{y} x^p y^q f(x,y) \quad p,q = 0,1,2……
\end{equation}

where f(x,y) is the pixel intensity value at the (x ,y)-coordinate.

Given the symmetrical arrangement of tooth sequences, the positional attribute is defined as the two-dimensional coordinates of the segmentation mask's center point subtracted by the absolute values of the coordinates of the overall panoramic radiograph's center point. And the centerline angle of the segmentation mask is determined by calculating the angle between the midline of the mask and the vertical direction. 

We design the glyph to visualize the multi-dimensional features of the segmentation mask(Figure \ref{component}(B)). The values of HU moments are encoded with a radial bar chart(Figure \ref{component}(B-b)). Within the glyph, we use the metaphor of a dashboard to encode the tooth’s two-dimensional coordinates(Figure \ref{component}(B-c,B-d)) and centerline angle(Figure \ref{component}(B-e)). To visually demonstrate the differences of features between the segmentation results and conventional training samples, we calculate the average value for each feature. Then we encode features close to the average value in gray, features significantly above the average value in blue, and features significantly below the average value in red(Figure \ref{component}(B-a)). The dental legend at the center of the glyph is populated with distinct colors according to the identified categories, facilitating a clearer observation of the tooth categorization(Figure \ref{component}(B-f)). Experts can effortlessly modify the assigned category labels by clicking on the dental legend.

\subsubsection{Feature Explorer Component}
For the reason that automatic segmentation algorithms rely on the matching between prior features and image characteristics, satisfactory segmentation results may not be achieved when there is prominent variation. In this section, we employ dimensionality reduction and mapping to obtain the standard range and distribution of multi-dimensional features for each category of teeth. By contrasting newly generated segmentation masks with the sample set distribution, we identify segmentation results that deviate from conventional patterns as which has a high probability of error.

Each point in the scatterplot view represents a tooth sample, with distinct colors indicating different categories. The manually annotated training and test sets are represented by points with higher transparency, while newly loaded tooth samples are differentiated by larger radii and lower transparency. To lay out the points in the scatterplot with respect to the feature similarities of the samples, we firstly employ the HU moments matrix to extract shape features from individual tooth slices, incorporating positional information within the original panoramic radiograph and centerline angular of the tooth to formulate a set of high-dimensional feature vectors. Then, we employ LDA\cite{blei2003latent} to project the vectors into a two-dimensional plane, generating a scatterplot, such that the samples share the similar features are closer. In general use, train samples are shown as solid circles, new loaded samples are shown as circles with black outlines and expert-specified samples are shown as crosses.

\begin{figure} 
\includegraphics[width=0.9\linewidth]{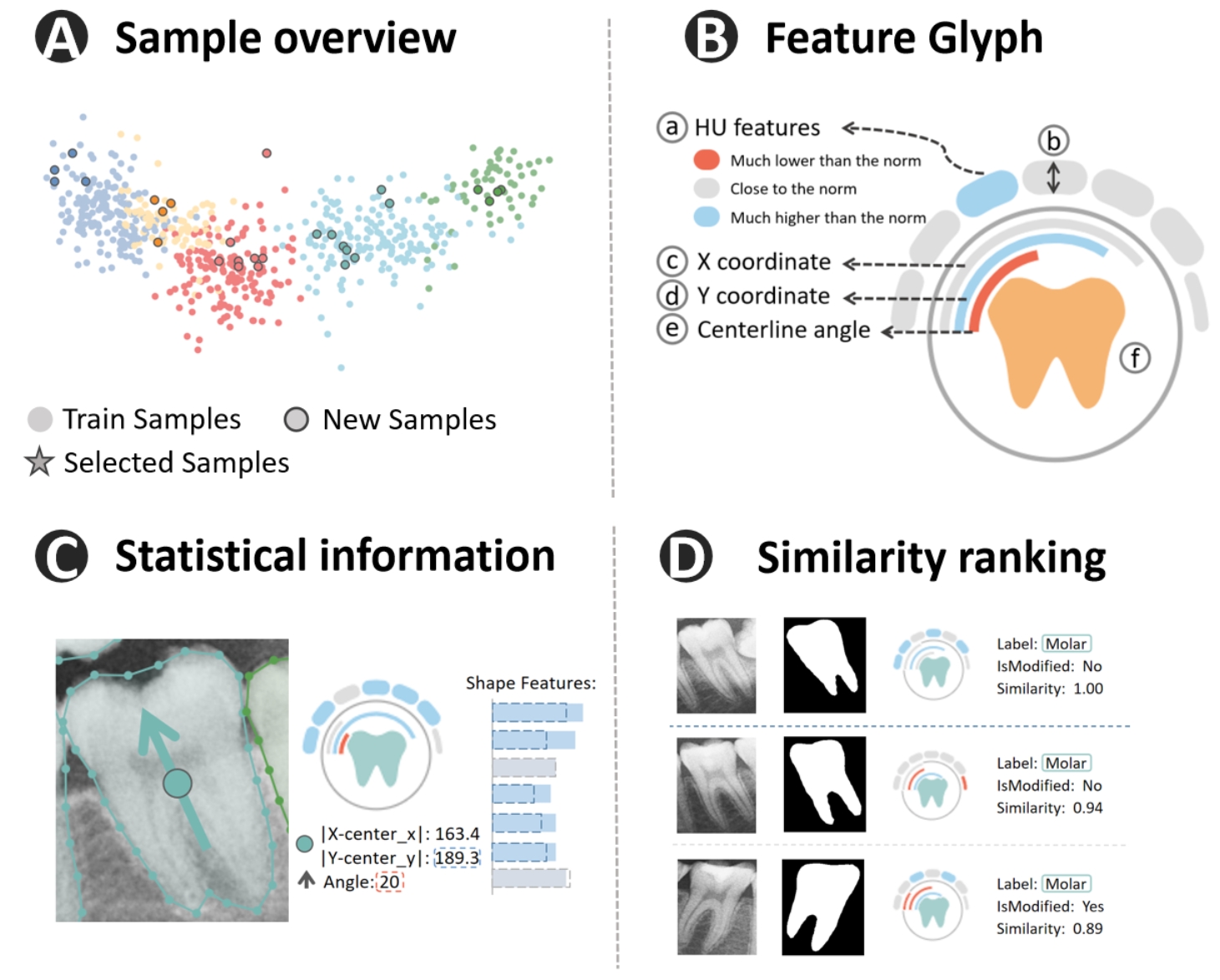}
\centering
\caption{ The extracted feature exploration component. The expert can start analysis from (A)the overview of the dataset. The (B)feature glyph, the (C)zoomed view and the (D)similarity view provide detailed information for feature exploration.} %最终文档中希望显示的图片标题
\label{component} %用于文内引用的标签
\end{figure}

The similarity view(Figure \ref{component}(D)) is designed to show historical labeled data with high similarity, providing an essential reference for whether masks are successfully identified or not. When the expert clicks on a tooth in the scatter plot, we calculate the historical labeled data adjacent to its projection position. Subsequently, the panorama slice map and the glyph will be presented in pairs in the similarity list and arranged in order of distance.

\subsection{Model Optimization}
The process begins with loading panoramic radiograph data for tooth segmentation. Firstly, the model output are projected in the scatterplot view, enabling experts to quickly discover the abnormal segmentation masks. The zoomed view and the reference view show different levels of detail, helping experts to do precise corrections manually. These expert corrections, functioning as high-quality labeled data, capture expert input on the accurate delineation of tooth boundaries. Once the necessary corrections are made, the projection view will update to show the new overview of the corrected results. Then the expert has the ability to choose several high-quality tooth samples and click ‘train’ in the control panel to feed the corrected high-quality labeled data back into the segmentation model. This step helps the model learn from the corrected data and improve its performance over time. The evaluation view provides a graphical representation of the optimization process, offering an intuitive insight into the model's performance throughout training. And the feedback loop continues as experts repeatedly load data, correct model outputs, and contribute to the ongoing refinement of the segmentation model.

\section{SYSTEM Interface}
We develop a set of interactions to integrate intelligent model and expert knowledge into the process of tooth segmentation. Initially, expert can gain automatic tooth segmentation by loading the panoramic radiograph in the control panel. Scatterplot view provides a compact overview for the standard range and distribution of multi-dimensional features for each tooth category. For more detailed features, expert can observe the glyph in zoomed view and similar samples in similarity view by clicking the corresponding scatter. By comparing the dissimilarity, experts can assess the consistency of feature distribution in the segmentation results with real structures. When the results of automatic segmentation deviate significantly from the normal range, further expert judgment and correction are necessary. Experts can make corrections to the delineation of tooth boundaries by clicking the anchor points on tooth. When the corrected segmentation is satisfactory, the scatterplot view will update to show the corrected results overview. Then experts have the ability to select high-quality labeled data which is considered expert feedback and contributes to improving the accuracy of the segmentation results. This iterative feedback loop helps improve the model's performance over time as it adapts to the corrections made by experts.

\section{EVALUATION}
We conducted two case studies and an expert study to demonstrate the effectiveness and usability of ViSTooth in tooth segmentation.

\subsection{Case Study}
\subsubsection{Case 1. Interactive Correction Insights}
\begin{figure*} 
\includegraphics[width=0.95\linewidth]{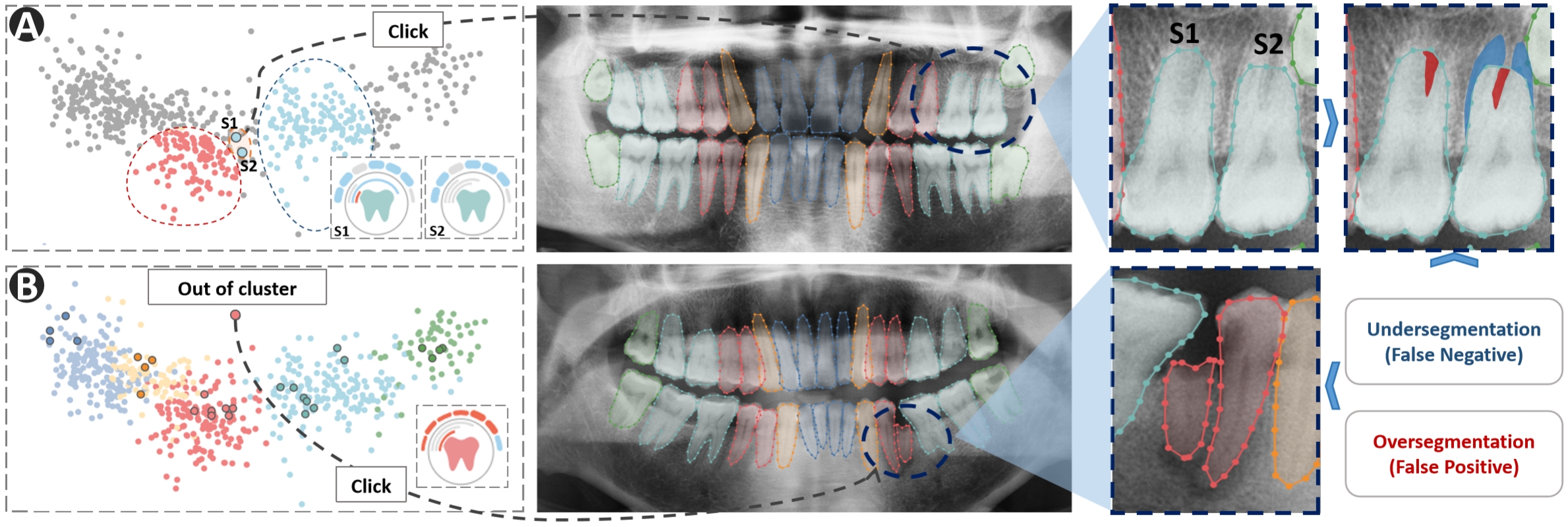}
\centering
\caption{ (A) displays the process of expert correction. (B) captures instances where individual variations lead to scattered outliers.} %最终文档中希望显示的图片标题
\label{case1} %用于文内引用的标签
\end{figure*}

We invited E1 to utilize VisTooth for detailed human-machine collaborative segmentation of teeth on 10 panoramic X-ray images, and asked him to follow the system's visual cues during the process. In the process of the segmentation task, all corrections and feedback were recorded. Initially, the segmentation model achieved an accuracy of 74.91\%, which was unsatisfactory. Thus E1 would like to use the system to inspect and refine the model outputs. Immediately, E1 identified some abnormal outliers from the scatter plot view(Figure \ref{case1}). He first clicked on an outlier to locate the tooth represented by it, concurrently the similarity view was updated to display detailed glyph and show samples similar in height to the selected sample. Upon observation, E1 found that one case of the outlier might be attributed to incomplete segmentation, leading to a significant separation between the mask and the regular distribution of that category. Figure \ref{case1}(A) illustrates how E1 corrected examples of teeth S1 and S2 by examining the morphology in the similarity view. Typically, the second molars have two roots, but due to the proximity of the pixel values between the root and the gingival tissue in the S1 and S2 regions, the model struggles to accurately differentiate tooth structure from other tissues. And the glyph in similarity view suggests that despite its mask features deviating from the second molar and resembling the first molar, its coordinate and angular features closely match those of the 2nd molars as predicted by the model. Subsequently, E1 attempted to adjust the contrast of the panoramic radiograph using the toolbar to enhance the differentiation between the target teeth and other structures, then manually adjusted the contour points to restore precise positioning. Furthermore, E1 also found that some cases characterized by individual differences could lead to outliers, as depicted in Figure \ref{case1}(B). Here, the patient exhibited incomplete tooth structures, significantly deviating from the training samples. Such situations bears the potential for erroneous segmentation, requiring manual assessment. E1 highly praised the glyph design, "Utilizing feature indicators as visual cues to help us detect segmentation anomalies for further manual correction is beneficial in the absence of ground truth for the newly loaded panoramic image."

\subsubsection{Case 2. Iterative Retraining Optimization}
\begin{figure} 
\includegraphics[width=0.9\linewidth]{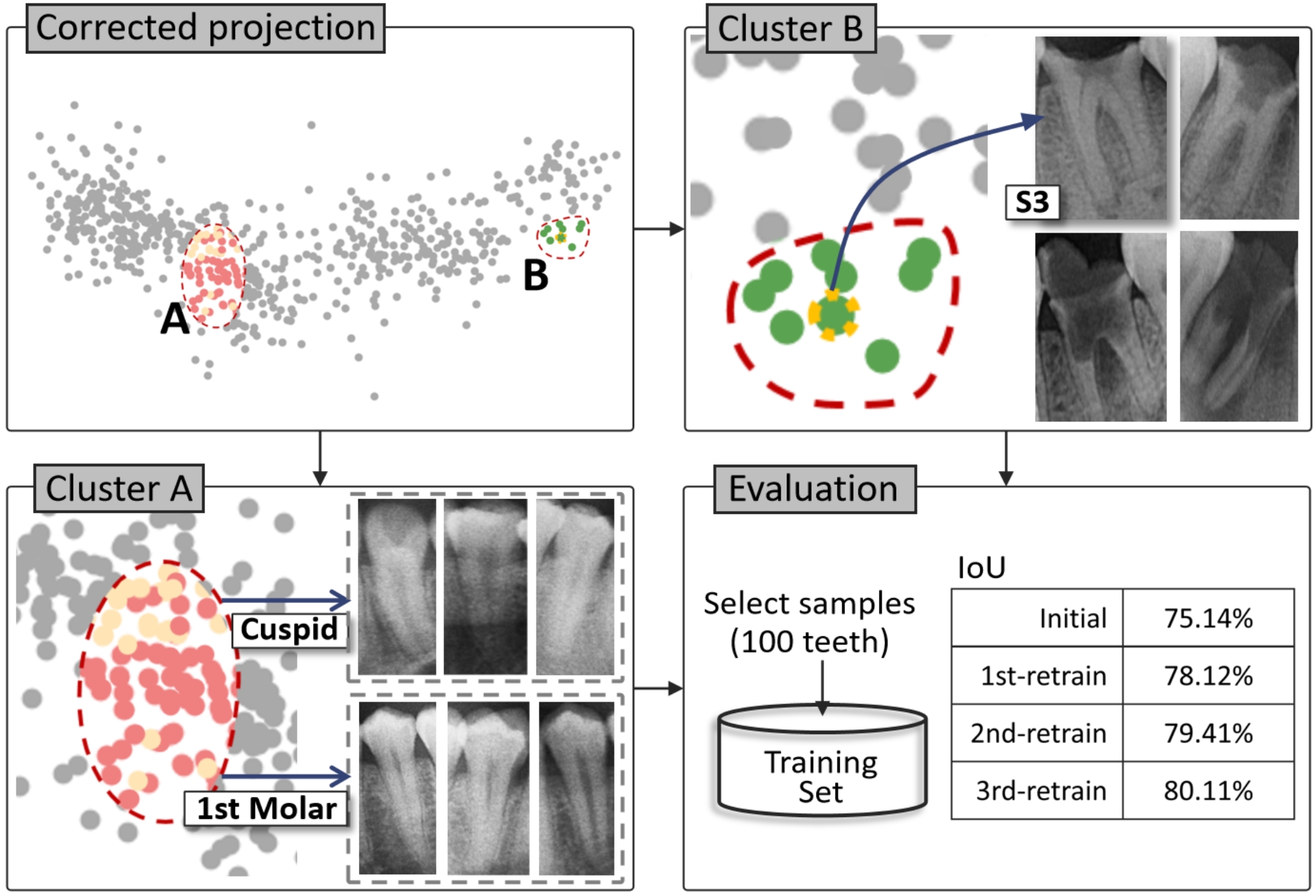}
\centering
\caption{Cluster A exhibited a mixed distribution pattern. Cluster B consistently differentiated into five major distribution patterns.} %最终文档中希望显示的图片标题
\label{case2} %用于文内引用的标签
\end{figure}

In the second case, we introduced more panoramic images. E2 was invited to perform batch panoramic segmentation and select samples for feedback to the model for retraining. Figure \ref{case2} shows the projection changes after manual correction by experts. The results indicate that cluster A exhibited a mixed distribution pattern during initial segmentation, and even after manual correction, clear differentiation was not achieved. E2 explained to us that the distinction between individual teeth is not particularly clear during actual reading, and there may be confusion between the cuspid and 1st molar labels (yellow and red) for the model. Therefore, E2 marked the mixed regions between these two patterns and added them to the training samples in the hope of strengthening the model's learning. Cluster B, on the other hand, consistently differentiated into five major distribution patterns. Figure \ref{case2} illustrates examples of S3 segmentation verification through Reference View examination. From this, we can see that S3 has a double-root structure similar to the reference view, but the significant crown loss deviates its shape features from the normal cluster. As the number of annotations increased, we observed the gradual aggregation of similar residual tooth clusters along the edge of S3. This feature is distinct from the training sample set and, therefore, E2 was eager to label it as a new sample to improve the model's recognition rate for residual teeth during segmentation. During the labeling process, E2 commented, "Using labeled samples to further enhance the model is very innovative. The improvement in initial segmentation accuracy means we can reduce manual correction." He also praised the visual attractiveness design of the projection view, noting that this distribution view effectively conveys the distribution pattern of segmentation masks and facilitates batch sample selection. Figure \ref{case2} shows the evaluation results of three retraining sessions, allowing experts to add 100 teeth slices with correction labels to the training set each time. The line graph illustrates the change in segmentation results before and after each retraining, demonstrating the effectiveness of our system in high-quality tooth segmentation. Initially, the IoU score was 75.14\%. After three rounds of training, significant improvement was observed, with the IoU score reaching 80.11\%.

\subsection{Expert Study}
\begin{figure} 
\includegraphics[width=1\linewidth]{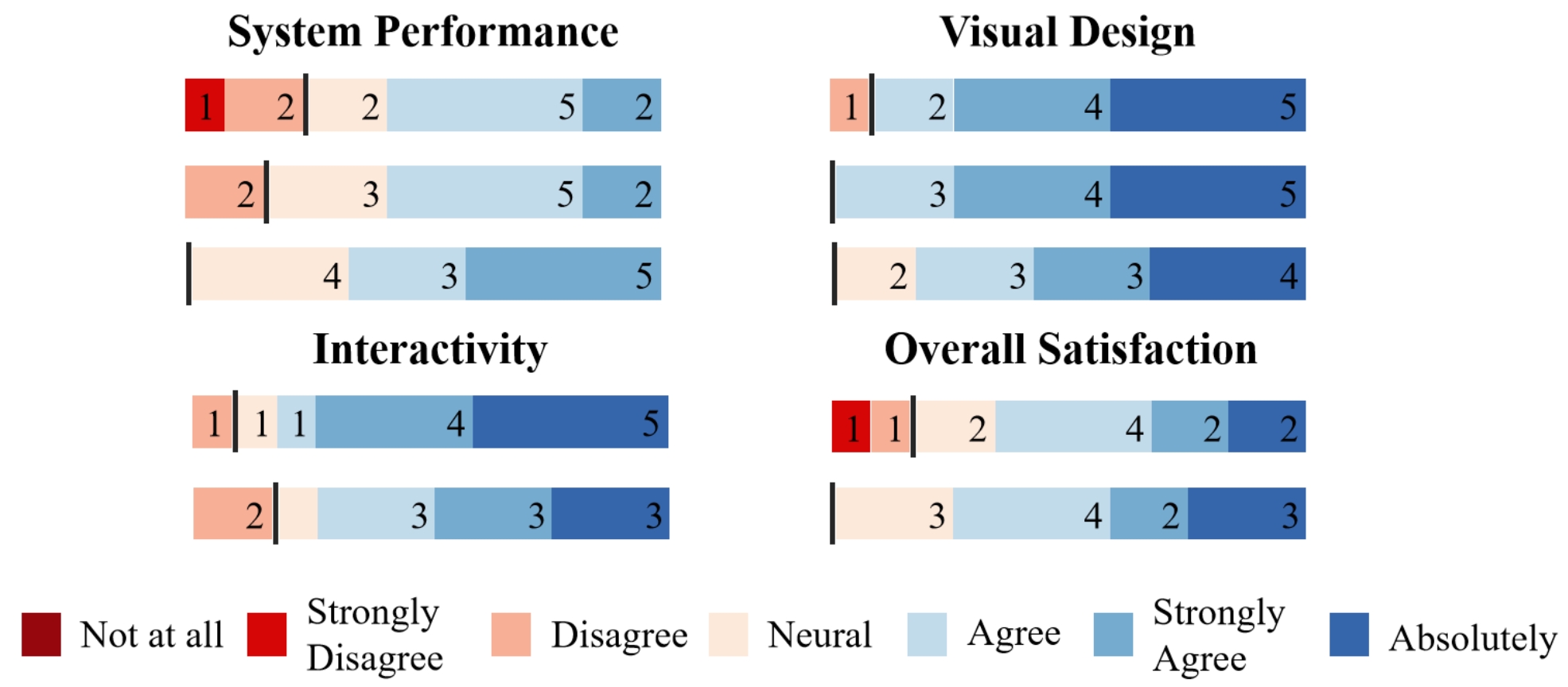}
\centering
\caption{ The feedback of the expert interviews.} %最终文档中希望显示的图片标题
\label{feadback} %用于文内引用的标签
\end{figure}

ViSTooth was designed to be an expressive and task efficient tool. To further evaluate the effectiveness of our system, we conducted an expert study involving 2 experts in dental examination and 10 graduate students (5 males and 5 females) majoring in Medicine. They were all trained to use our system until they were familiar with the workflow and proficient in utilizing the system. Thereafter, they were tasked with the segmentation of 60 panoramic radiographs.  During the process, we recorded their comments and the interactions. Further, we formulated a set of questions, which are closely related to the analytical tasks outlined in Section 3. The questionnaire is displayed in Table \ref{tabel2}, and participants’ responses can be observed in Figure \ref{feadback}. Here are some key findings from the analysis:

\textbf{System Performance.} The majority of participants expressed satisfaction with the accuracy and speed of the preliminary segmentation performed by the AutoML model. E1 commented,”The proposed model can effectively support preliminary segmentation, which alleviates laborious and time-consuming manual detection.” Statistical analysis showed that 75\% of participants rated the accuracy as satisfactory, while 83\% were satisfied with the speed. Analysis of the collected metrics indicated that they effectively reflected the quality of the segmentation results, with 80\% of participants agreeing with this statement.

\textbf{Visual Design.} Over 90\% of participants found the interface design to be intuitive and easy to understand, highlighting the effectiveness of the visual design in facilitating user interaction. An overwhelming majority (over 95\%) of participants agreed that the color choices and graphical elements in the system contributed to detecting segmentation anomalies, underscoring the importance of visual cues in the analysis process. E2 remarked, “The visual design of ViSTooth greatly facilitates the interpretation of segmentation results, making it easier to identify abnormalities.” 

\textbf{Interactivity.} The interactive features designed for digging deeper and gaining more insights into the segmentation results were well-received, with 83\% of participants expressing satisfaction with this aspect. Similarly, the interactive feature design for adjusting segmentation results garnered positive feedback, with 75\% of participants reporting satisfaction. One graduate student noted, “The interactive features provide flexibility and control, allowing for fine-tuning of segmentation results according to individual preferences.”

\textbf{Overall Satisfaction.} A significant portion of participants found ViSTooth to be easy to use, indicating high overall satisfaction with the system's usability. Impressively, 75\% of participants expressed willingness to continue using ViSTooth in their future clinical practice, reflecting a strong endorsement of the system's utility and effectiveness. However, a minority of participants expressed concerns about mastering ViSTooth's advanced features, suggesting the need for additional training resources or user guides.

These statistical findings provide robust evidence supporting the positive reception of ViSTooth among users, affirming its effectiveness as an expressive and task-efficient tool for panoramic radiograph segmentation.

\begin{table}
\label{tabel2}
\caption{The questionnaire consists of four parts: the system performance (Q1-3), the visual design (Q4-6), the interactivity (Q7-8), and the overall satisfaction (Q9-10).}
\renewcommand\arraystretch{1.5}
\begin{tabular}{l|p{7cm}}
\hline
Q1  & I am satisfied with the accuracy of preliminary tooth segmentation by the AutoML model.                                              \\ \cline{2-2} 
Q2  & I am satisfied with the speed of tooth segmentation by the AutoML model.                                                             \\ \cline{2-2} 
Q3  & The metrics proposed can reflect the quality of the tooth segmentation results.                                                       \\ \hline
Q4  & The interface design is intuitive and easy to understand.                                                                      \\ \cline{2-2} 
Q5  & The color choices and graphical elements in the system contribute to detecting segmentation anomalies.                         \\ \cline{2-2} 
Q6  & The layout of the system's interface contributes to my ease of understanding and using its features.                           \\ \hline
Q7  & I am satisfied with the interactive feature design for digging deeper and gaining more insights into the tooth segmentation results.    \\ \cline{2-2} 
Q8  & I am satisfied with the provided tools and controls for adjusting the tooth segmentation results.                                     \\ \hline
Q9  & ViSTooth is easy to use.                                                                                                       \\ \cline{2-2} 
Q10 & I am willing to continue using this system in clinical practice.                                                                  \\ \hline
\end{tabular}
\end{table}

\section{Discussion}
\textbf{Model performance.} Automation of tooth segmentation is considered the first and foundational step in the development of AI systems for adjuvant therapy in dentistry. Therefore, this first step should be as accurate as possible. We focus on the revolutionary impact of Large Language Models (LLMs), such as ChatGPT\cite{chatgpt}, SAM\cite{kirillov2023segment}, has permeated various industries. We believe that the advanced language understanding, contextual interpretation and more nuanced feature recognition abilities of LLMs can enhance the segmentation process.

\textbf{Feature indicators.} Automatic evaluation is crucial in efficiently guiding experts to improve segmentation quality. Starting from the common characteristics of teeth, this paper extracts tooth angles, positions, and shapes to screen out results with higher error probabilities. However, personalized differences among teeth, such as the proximity between adjacent teeth, treatment marks, and developmental stages, can affect this assessment. Therefore, in future work, we plan to explore more extensively how to utilize richer features to characterize the quality of segmentation results, such as internal density distribution, texture features, and edge features of teeth.

\textbf{Automated Diagnosis} Tooth segmentation is the most widely used processing technique to analyze panoramic radiographs. With precisely segmented tooth structures, further applications can be developed in computer-aided dental diseases, such as diagnosis, tooth alignment assessment, orthodontic optimization, etc. This work forms the basis of our further developments of AI-driven tools for precise and automated diagnosis of various dental diseases\cite{leite2020radiomics, dayi2023novel, sivari2023deep}. By leveraging these developments, we hope to foster efficiency and accuracy in dental healthcare delivery.

\section{CONCLUSION}
In this paper, we present ViSTooth for accurate tooth segmentation through human-machine collaboration. Based on domain expertise, the model in ViSTooth automatically preliminary tooth segmentation. Then the visual interface provides various supporting information to help experts to learn the segmentation results and detect anomalies. Rich human computer interactions are integrated to enable higher quality corrected data and iterative optimization of the segmentation model. Two case studies and an expert study highlight the effectiveness of our tool in streamlining the tooth segmentation process and minimizing the manual effort required for accurate results. In the future work, we hope to improve the performance of automatic segmentation to further reduce the effort of manual correction, leverage richer features for automatic evaluation, as well as integrate tooth segmentation into disease diagnosis and treatment applications.
%%%%%%%%%%%%%%%%%%%%%%%%%%%%%%%%%%%%%%%%%%%%%%%%%%%%%

% \bibliographystyle{abbrv}
% \bibliographystyle{abbrv-doi}
\bibliographystyle{unsrt} %%%

\bibliography{template}

\begin{thebibliography}{10}

\bibitem{jcm9072313}
Vanessa Machado, Luís Proença, Mariana Morgado, José~João Mendes, and João Botelho.
\newblock Accuracy of panoramic radiograph for diagnosing periodontitis comparing to clinical examination.
\newblock {\em Journal of Clinical Medicine}, 9(7), 2020.

\bibitem{diagnostics14050497}
Shaofeng Wang, Shuang Liang, Qiao Chang, Li~Zhang, Beiwen Gong, Yuxing Bai, Feifei Zuo, Yajie Wang, Xianju Xie, and Yu~Gu.
\newblock Stsn-net: Simultaneous tooth segmentation and numbering method in crowded environments with deep learning.
\newblock {\em Diagnostics}, 14(5), 2024.

\bibitem{2018Automated}
Jie Yang, Yuchen Xie, Lin Liu, Bin Xia, Zhanqiang Cao, and Chuanbin Guo.
\newblock Automated dental image analysis by deep learning on small dataset.
\newblock pages 492--497, 2018.

\bibitem{ACompre2023}
A comprehensive review of recent advances in artificial intelligence for dentistry e-health.
\newblock 2023.

\bibitem{Develop2023}
Developing deep learning methods for classification of teeth in dental panoramic radiography.
\newblock 2023.

\bibitem{Indraswari2015}
Rarasmaya Indraswari, Agus~Zainal Arifin, Dini~Adni Navastara, and Naser Jawas.
\newblock Teeth segmentation on dental panoramic radiographs using decimation-free directional filter bank thresholding and multistage adaptive thresholding.
\newblock In {\em 2015 International Conference on Information \& Communication Technology and Systems (ICTS)}, pages 49--54, 2015.

\bibitem{Mohamed2014}
Muhamad~Rizal Mohamed~razali, Nazatul~Sabariah Ahmad, Zulkifly Mohd~Zaki, and Waidah Ismail.
\newblock Region of adaptive threshold segmentation between mean, median and otsu threshold for dental age assessment.
\newblock pages 353--356, 2014.

\bibitem{razali2014}
Muhamad Rizal~Mohamed Razali, Nazatul~Sabariah Ahmad, Rozita Hassan, Zulkifly~Mohd Zaki, and Waidah Ismail.
\newblock Sobel and canny edges segmentations for the dental age assessment.
\newblock In {\em 2014 International Conference on Computer Assisted System in Health}, pages 62--66. IEEE, 2014.

\bibitem{2012fuzzy}
N~Senthilkumaran.
\newblock Fuzzy logic approach to edge detection for dental x-ray image segmentation.
\newblock {\em International Journal of Computer Science and Information Technologies}, 3(5):5236--5238, 2012.

\bibitem{li2022semantic}
Pengcheng Li, Yang Liu, Zhiming Cui, Feng Yang, Yue Zhao, Chunfeng Lian, and Chenqiang Gao.
\newblock Semantic graph attention with explicit anatomical association modeling for tooth segmentation from cbct images.
\newblock {\em IEEE Transactions on Medical Imaging}, 41(11):3116--3127, 2022.

\bibitem{SILVA201815}
Gil Silva, Luciano Oliveira, and Matheus Pithon.
\newblock Automatic segmenting teeth in x-ray images: Trends, a novel data set, benchmarking and future perspectives.
\newblock {\em Expert Systems with Applications}, 107:15--31, 2018.

\bibitem{Unet2023}
Senbao Hou, Tao Zhou, Yuncan Liu, Pei Dang, Huiling Lu, and Hongbin Shi.
\newblock Teeth u-net: A segmentation model of dental panoramic x-ray images for context semantics and contrast enhancement.
\newblock {\em Computers in Biology and Medicine}, 152:106296, 2023.

\bibitem{Koch2019}
Thorbjørn~Louring Koch, Mathias Perslev, Christian Igel, and Sami~Sebastian Brandt.
\newblock Accurate segmentation of dental panoramic radiographs with u-nets.
\newblock pages 15--19, 2019.

\bibitem{chen2019deep}
Hu~Chen, Kailai Zhang, Peijun Lyu, Hong Li, Ludan Zhang, Ji~Wu, and Chin-Hui Lee.
\newblock A deep learning approach to automatic teeth detection and numbering based on object detection in dental periapical films.
\newblock {\em Scientific reports}, 9(1):3840, 2019.

\bibitem{kim2020automatic}
Changgyun Kim, Donghyun Kim, HoGul Jeong, Suk-Ja Yoon, and Sekyoung Youm.
\newblock Automatic tooth detection and numbering using a combination of a cnn and heuristic algorithm.
\newblock {\em Applied Sciences}, 10(16):5624, 2020.

\bibitem{2020study}
Bernardo Silva, La{\'\i}s Pinheiro, Luciano Oliveira, and Matheus Pithon.
\newblock A study on tooth segmentation and numbering using end-to-end deep neural networks.
\newblock pages 164--171, 2020.

\bibitem{Schwendicke2020}
Krois J.~Artificial Schwendicke~F, Samek~W.
\newblock Intelligence in dentistry: Chances and challenges.
\newblock {\em Journal of Dental Research}, pages 769--774, 2020.

\bibitem{leite2021artificial}
Andr{\'e}~Ferreira Leite, Adriaan~Van Gerven, Holger Willems, Thomas Beznik, Pierre Lahoud, Hugo Ga{\^e}ta-Araujo, Myrthel Vranckx, and Reinhilde Jacobs.
\newblock Artificial intelligence-driven novel tool for tooth detection and segmentation on panoramic radiographs.
\newblock {\em Clinical oral investigations}, 25:2257--2267, 2021.

\bibitem{maskrcnn2017}
Kaiming He, Georgia Gkioxari, Piotr Dollár, and Ross Girshick.
\newblock Mask r-cnn.
\newblock pages 2980--2988, 2017.

\bibitem{DBLP2021}
Inkyu Shin, Dong{-}Jin Kim, Jae{-}Won Cho, Sanghyun Woo, KwanYong Park, and In~So Kweon.
\newblock Labor: Labeling only if required for domain adaptive semantic segmentation.
\newblock {\em CoRR}, abs/2108.05570, 2021.

\bibitem{Modi2011}
Chintan~K. Modi and Nirav~P. Desai.
\newblock A simple and novel algorithm for automatic selection of roi for dental radiograph segmentation.
\newblock In {\em 2011 24th Canadian Conference on Electrical and Computer Engineering(CCECE)}, pages 000504--000507, 2011.

\bibitem{Alsmadi2018}
Mutasem~K Alsmadi.
\newblock A hybrid fuzzy c-means and neutrosophic for jaw lesions segmentation.
\newblock {\em Ain Shams Engineering Journal}, 9(4):697--706, 2018.

\bibitem{Hasan2016}
Mosaddik Hasan, Waidah~Binti Ismail, Rozita Hassan, and Atsuo Yoshitaka.
\newblock Automatic segmentation of jaw from panoramic dental x-ray images using gvf snakes.
\newblock {\em 2016 World Automation Congress (WAC)}, pages 1--6, 2016.

\bibitem{Li2012}
Hui Li, Guoxia Sun, Huiqiang Sun, and W.~Liu.
\newblock Watershed algorithm based on morphology for dental x-ray images segmentation.
\newblock {\em 2012 IEEE 11th International Conference on Signal Processing}, 2:877--880, 2012.

\bibitem{Fariza2019}
Arna Fariza, Agus~Zainal Arifin, Eha~Renwi Astuti, and Takio Kurita.
\newblock Segmenting tooth components in dental x-ray images using gaussian kernel- based conditional spatial fuzzy c-means clustering algorithm.
\newblock {\em International Journal of Intelligent Engineering and Systems}, 2019.

\bibitem{jader2018}
Gil Jader, Jefferson Fontineli, Marco Ruiz, Kalyf Abdalla, Matheus Pithon, and Luciano Oliveira.
\newblock Deep instance segmentation of teeth in panoramic x-ray images.
\newblock pages 400--407, 2018.

\bibitem{ALmalki2023}
A.~Almalki and L.~Latecki.
\newblock Self-supervised learning with masked image modeling for teeth numbering, detection of dental restorations, and instance segmentation in dental panoramic radiographs.
\newblock pages 5583--5592, jan 2023.

\bibitem{Zhang2018AnET}
Kailai Zhang, Ji~Wu, Hu~Chen, and Peijun Lyu.
\newblock An effective teeth recognition method using label tree with cascade network structure.
\newblock {\em Computerized medical imaging and graphics : the official journal of the Computerized Medical Imaging Society}, 68:61--70, 2018.

\bibitem{helli2022}
Serdar Helli and Anda{\c{c}} Hamamc{\i}.
\newblock Tooth instance segmentation on panoramic dental radiographs using u-nets and morphological processing.
\newblock {\em D{\"u}zce {\"U}niversitesi Bilim ve Teknoloji Dergisi}, 10(1):39--50, 2022.

\bibitem{tuzoff2019tooth}
Dmitry~V Tuzoff, Lyudmila~N Tuzova, Michael~M Bornstein, Alexey~S Krasnov, Max~A Kharchenko, Sergey~I Nikolenko, Mikhail~M Sveshnikov, and Georgiy~B Bednenko.
\newblock Tooth detection and numbering in panoramic radiographs using convolutional neural networks.
\newblock {\em Dentomaxillofacial Radiology}, 48(4):20180051, 2019.

\bibitem{bhat2023}
Suvarna Bhat, Gajanan~K Birajdar, and Mukesh~D Patil.
\newblock A comprehensive survey of deep learning algorithms and applications in dental radiograph analysis.
\newblock {\em Healthcare Analytics}, page 100282, 2023.

\bibitem{VisAI2023}
Xumeng Wang, Ziliang Wu, Wenqi Huang, Wei Yating, Zhaosong Huang, Mingliang Xu, and Wei Chen.
\newblock Vis+ai: integrating visualization with artificial intelligence for efficient data analysis.
\newblock {\em Frontiers of Computer Science}, 17, 06 2023.

\bibitem{DAI2018}
Wenjing Dai, Meng Wang, Zhibin Niu, and Jiawan Zhang.
\newblock Chart decoder: Generating textual and numeric information from chart images automatically.
\newblock {\em Journal of Visual Languages \& Computing}, 48:101--109, 2018.

\bibitem{Ood}
Changjian Chen, Jun Yuan, Yafeng Lu, Yang Liu, Hang Su, Songtao Yuan, and Shixia Liu.
\newblock Oodanalyzer: Interactive analysis of out-of-distribution samples.
\newblock {\em IEEE Transactions on Visualization and Computer Graphics}, 27(7):3335–3349, jul 2021.

\bibitem{kandel2012}
Sean Kandel, Ravi Parikh, Andreas Paepcke, Joseph~M Hellerstein, and Jeffrey Heer.
\newblock Profiler: Integrated statistical analysis and visualization for data quality assessment.
\newblock pages 547--554, 2012.

\bibitem{liu2017analyzing}
Mengchen Liu, Jiaxin Shi, Kelei Cao, Jun Zhu, and Shixia Liu.
\newblock Analyzing the training processes of deep generative models.
\newblock {\em IEEE transactions on visualization and computer graphics}, 24(1):77--87, 2017.

\bibitem{cao2020analyzing}
Kelei Cao, Mengchen Liu, Hang Su, Jing Wu, Jun Zhu, and Shixia Liu.
\newblock Analyzing the noise robustness of deep neural networks.
\newblock {\em IEEE Transactions on Visualization and Computer Graphics}, 27(7):3289--3304, 2020.

\bibitem{wang2021}
Zijie~J. Wang, Robert Turko, Omar Shaikh, Haekyu Park, Nilaksh Das, Fred Hohman, Minsuk Kahng, and Duen~Horng Polo~Chau.
\newblock Cnn explainer: Learning convolutional neural networks with interactive visualization.
\newblock {\em IEEE Transactions on Visualization and Computer Graphics}, 27(2):1396--1406, 2021.

\bibitem{Ma2015}
Aravindh Mahendran and Andrea Vedaldi.
\newblock Understanding deep image representations by inverting them.
\newblock pages 5188--5196, 2015.

\bibitem{Sel2017}
Ramprasaath~R. Selvaraju, Michael Cogswell, Abhishek Das, Ramakrishna Vedantam, Devi Parikh, and Dhruv Batra.
\newblock Grad-cam: Visual explanations from deep networks via gradient-based localization.
\newblock pages 618--626, 2017.

\bibitem{chen2023unified}
Changjian Chen, Yukai Guo, Fengyuan Tian, Shilong Liu, Weikai Yang, Zhaowei Wang, Jing Wu, Hang Su, Hanspeter Pfister, and Shixia Liu.
\newblock A unified interactive model evaluation for classification, object detection, and instance segmentation in computer vision.
\newblock {\em IEEE Transactions on Visualization and Computer Graphics}, 2023.

\bibitem{ahn2019fairsight}
Yongsu Ahn and Yu-Ru Lin.
\newblock Fairsight: Visual analytics for fairness in decision making.
\newblock {\em IEEE transactions on visualization and computer graphics}, 26(1):1086--1095, 2019.

\bibitem{pixelfeature2023}
Maaz Ansari, Surendra Bhosale, and Archana Choudhary.
\newblock Semantic segmentation using convolutional neural networks.
\newblock 10:31--34, 06 2023.

\bibitem{he2016deep}
Kaiming He, Xiangyu Zhang, Shaoqing Ren, and Jian Sun.
\newblock Deep residual learning for image recognition.
\newblock pages 770--778, 2016.

\bibitem{VGG2014}
Karen Simonyan and Andrew Zisserman.
\newblock Very deep convolutional networks for large-scale image recognition.
\newblock {\em arXiv preprint arXiv:1409.1556}, 2014.

\bibitem{hu1962visual}
Ming-Kuei Hu.
\newblock Visual pattern recognition by moment invariants.
\newblock {\em IRE transactions on information theory}, 8(2):179--187, 1962.

\bibitem{HUap2020}
Frederik~J.S. Doerr and Alastair~J. Florence.
\newblock A micro-xrt image analysis and machine learning methodology for the characterisation of multi-particulate capsule formulations.
\newblock {\em International Journal of Pharmaceutics: X}, 2:100041, 2020.

\bibitem{blei2003latent}
David~M Blei, Andrew~Y Ng, and Michael~I Jordan.
\newblock Latent dirichlet allocation.
\newblock {\em Journal of machine Learning research}, 3(Jan):993--1022, 2003.

\bibitem{chatgpt}
OpenAI.
\newblock Introducing chatgpt.

\bibitem{kirillov2023segment}
Alexander Kirillov, Eric Mintun, Nikhila Ravi, Hanzi Mao, Chloe Rolland, Laura Gustafson, Tete Xiao, Spencer Whitehead, Alexander~C Berg, Wan-Yen Lo, et~al.
\newblock Segment anything.
\newblock pages 4015--4026, 2023.

\bibitem{leite2020radiomics}
André~Ferreira Leite, Karla de~Faria Vasconcelos, Holger Willems, and Reinhilde Jacobs.
\newblock Radiomics and machine learning in oral healthcare.
\newblock {\em PROTEOMICS--Clinical Applications}, 14(3):1900040, 2020.

\bibitem{dayi2023novel}
Burak Day{\i}, H{\"u}seyin {\"U}zen, {\.I}pek~Bal{\i}k{\c{c}}{\i} {\c{C}}i{\c{c}}ek, and {\c{S}}uayip~Burak Duman.
\newblock A novel deep learning-based approach for segmentation of different type caries lesions on panoramic radiographs.
\newblock {\em Diagnostics}, 13(2):202, 2023.

\bibitem{sivari2023deep}
Esra Sivari, Guler~Burcu Senirkentli, Erkan Bostanci, Mehmet~Serdar Guzel, Koray Acici, and Tunc Asuroglu.
\newblock Deep learning in diagnosis of dental anomalies and diseases: A systematic review.
\newblock {\em Diagnostics}, 13(15):2512, 2023.

\end{thebibliography}
\end{document}